\documentclass{article}
\usepackage{amssymb}
\usepackage{amsfonts}
\usepackage{epsfig}

\newcommand{\be}{\begin{equation}}
\newcommand{\ee}{\end{equation}}

\begin{document}

\title{Cavity with an embedded polarized film: an adapted spectral approach}
\author{J.-G.~Caputo$^{1~*)}$, E. V.~Kazantseva$^1$, L. Loukitch$^1$ and A.I.~Maimistov$^2$}
\maketitle


\begin{center}
{\normalsize \textit{1) Laboratoire de Math\'ematiques, INSA de Rouen,}}\\[%
0pt]
{\normalsize \textit{B.P. 8, 76131 Mont-Saint-Aignan cedex, France}} \\[0pt]
{\normalsize \textit{\phantom{1)} E-mail: caputo@insa-rouen.fr}} \\[0pt]
{\normalsize \textit{B.P. 8, 76131 Mont-Saint-Aignan cedex, France}} \\[0pt]
{\normalsize \textit{\phantom{1)} E-mail: murkamars@hotmail.com }}\\[0pt]
{\normalsize \textit{2) Department of Solid State Physics,}} \\[0pt]
{\normalsize \textit{Moscow Engineering Physics Institute,}}\\[0pt]
{\normalsize \textit{Kashirskoe sh. 31, Moscow, 115409 Russia}}\\[0pt]
{\normalsize \textit{\phantom{1)} E-mail: amaimistov@hotmail.com}}
\end{center}



\begin{abstract}
We consider the modes of the electric field of a cavity where there
is an embedded polarized dielectric film. The model consists in
the Maxwell equations coupled to a Duffing oscillator for
the film which we assume infinitely thin. We derive the normal
modes of the system and show that they are orthogonal with a special
scalar product which we introduce. These modes are well suited to describe
the system even for a film of finite thickness. By acting on the film
we demonstrate switching from one cavity mode to another.
Since the system is linear, little energy is needed for this 
conversion. Moreover the amplitude equations describe very well
this complex system under different perturbations (damping, forcing and
nonlinearity) with very few modes. 
These results
are very general and can be applied to different situations like
for an atom in a cavity or a Josephson junction in a capacitor and
this could be very useful for many nano-physics applications.
\end{abstract}

\section{Introduction}

There are many reasons to couple an oscillator to a cavity.
One example is a laser built using a
Fabry-Perot resonator enclosing an active medium
which can be modeled as a two level atom \cite{nm92,Walther}.
The cavity can also be used to synchonize oscillators
\cite{voabl03} as for an array of Josephson junctions.
For window Josephson junctions, used as microwave generators, the
Josephson junction collects all the energy in one of the cavity modes
\cite{cl07}.
The cavity can also be used as a thermostat to cool
the oscillator as described in \cite{hhgsr97} where
an optical cavity is used to cool an atom. In another example
coupling an atom or quantum oscillator to a resonator can significantly
change its transport properties \cite{sf05}.

In all the systems described above we have a localized oscillator
coupled to a resonator. In addition the size of the oscillator can
often be neglected.
This situation can be represented by a thin film model
\cite{ry82,ry87}. Such a thin film model was generalized by 
taking into account the local field effects (dipole-dipole interaction)
\cite{BBE86,bzmt89,bmtz91}. Intrinsic optical bistability is the
main result of this generalization. Thin films containing
three-level atoms \cite{em01}, two-photon resonant atoms
\cite{bem99,em99}, inhomogeneously broadened two-level
\cite{Elyutin07} atoms and  two-level atoms with permanent dipole
\cite{Elyutin07a} represent the different generalizations of the
model. The coherent responses of resonant atoms of a thin film to short
optical pulse excitation were considered in \cite{Elyutin07}. It was
shown that for a certain intensity the incident pulse generates sharp
spikes in the transmitted radiation. Photon echo in
the form of multiple responses to a double or triple pulse
excitation was predicted also in this paper. The coherent reflection
from a thin film as superradiation was studied in
\cite{bt88,Malyshev00,lst90}.

Recently we used the thin film model to describe switching
phenomena in ferroelectrics \cite{ckm06}. The behavior of the new
artificial materials -- metamaterials -- under action of
electromagnetic pulses could be also described by this model
\cite{maga07}.
As we see the thin film model is extremely fruitful, the investigation of
the behavior of the thin film embedded inside the cavity is a very
attractive problem.
When the model of thin film is explored, the problem of the
matter-field interaction reduces to the one of an electromagnetic
cavity with an embedded (linear or nonlinear) oscillator. Frequently
the nonlinear problem is analyzed using coupled mode
equations. In this approach the relevant values are the amplitudes 
of each of the linear modes. The nonlinear partial differential
equation of motion
is reduced to a system of coupled ordinary differential
equations for the amplitudes of the linear
modes. A suitable choice of the linear modes allows to get an
effective description of the original problem.

We consider here this simplest and most general situation, first for
small energy for which the problem is linear. In this case, the
system film/cavity has normal modes which we characterize. These 
describe correctly the evolution of the system as opposed to 
standard Fourier modes or even other simpler normal modes.
These adapted normal modes are
orthogonal with respect to an adapted scalar product. Using these
modes we can define simply the state of the system and obtain mode
conversion by acting on the film sub-system. In this paper we
propose the mode conversion mechanism for a linear system which can
be used for different applications. In the nonlinear case, for
medium amplitudes,
we show that a few modes are sufficient to describe the evolution of the system. \\
After introducing the model in section 2, we consider the linear limit in
section 3 and derive the normal modes of the system. The special
scalar product is derived in section 4. In sections 5 and 6 we use the
normal modes to define the state of the (linear) system and show
mode conversion when driving and damping the film. We also
describe the general nonlinear case and we conclude in section 7.

\section{The thin film model}

We consider a one dimensional model of the electromagnetic radiation
interacting with a polarized dielectric film inside a cavity. The
film is placed at the distance $ x_a$ inside the cavity having the
length $l$. The Lagrangian density for the electromagnetic field,
the film medium and their coupling is the following \cite{ckm06}:
\begin{equation}
\mathcal{L}= \frac{a_{t}^{2}}{2} -\frac{a_{x}^{2}}{2} +\delta \left(
x-x_a\right)\left( {\frac{q_{t}^{2}}{2}- m\frac{q^{2}}{2} -
\frac{q^{4}}{4}-\alpha qa_{t}}\right). \label{lag_ap}
\end{equation}
Here $a$ is the analog of vector potential and $q$ is the medium polarization, $
\alpha$ is a coupling constant. The last term in Lagrangian describes
the coupling between $a$ and $q$. The dielectric medium can be
ferroelectric ($m=-1$) with two polarizations or paraelectric $m=1$.
The Hamiltonian of the system is
\begin{equation}
\mathcal{H}=a_{t}\frac{\partial \mathcal{L}}{\partial
a_{t}}+q_{t}\frac{
\partial \mathcal{L}}{\partial q_{t}}-\mathcal{L},  \label{ham_ap1}
\end{equation}
which gives
\begin{equation}
\mathcal{H}= \frac{a_{t}^{2}}{2} +\frac{a_{x}^{2}}{2} +\delta \left(
x-x_a\right) \left( {\frac{q_{t}^{2}}{2}+ m\frac{q^{2}}{2} +
\frac{q^{4}}{4}}\right), \label{ham_ap}
\end{equation}
The variation of the action functional yields the Euler-Lagrange equations
for $a$ and $q$
\begin{equation}
\frac{\partial \mathcal{L}}{\partial a}=\frac{d}{dt}\frac{\partial
\mathcal{L }}{\partial a_{t}}+\frac{d}{dx}\frac{\partial
\mathcal{L}}{\partial a_{x}} \label{el1}
\end{equation}
\begin{equation}
\frac{\partial \mathcal{L}}{\partial q}=\frac{d}{dt}\frac{\partial
\mathcal{L }}{\partial q_{t}}+\frac{d}{dx}\frac{\partial
\mathcal{L}}{\partial q_{x}} \label{el2}
\end{equation}
which reduce to
\begin{eqnarray}
a_{tt}-a_{xx} &=&\alpha \delta \left( x-x_a\right)q_{t}  \label{aq} \\
q_{tt}+mq+ q^{3} &=-&\alpha a_{t}
\end{eqnarray}
The equations for the electric field $e=-a_{t}$ and medium variable can then
be obtained
\begin{eqnarray}
e_{tt}-e_{xx} &=&-\alpha \delta \left( x-x_a\right)q_{tt},  \label{eq1} \\
q_{tt}+mq+ q^{3} &=&\alpha e(x_a),
\end{eqnarray}
where the coupling between the fields $e$ and $q$ only occurs in the medium
at $x=x_a$.

In a recent article \cite{ckm06}, we considered with this model the
interaction of a thin dielectric film with an electromagnetic pulse.
We studied both the case of a ferroelectric and paraelectric film.
For the ferroelectric film we showed that the polarization can be
switched by an incoming pulse and studied this phenomenon. Here we
will assume that the film is embedded in a cavity and we will
study how cavity modes can be controlled by the film. Specifically
we will assume Dirichlet boundary conditions for the field.

For small amplitudes of the field, it is natural to neglect the
nonlinear response of the film. Note however that the ferroelectric
film and paraelectric film have different natural frequencies
corresponding to different stationary points. For the paraelectric
case, there is only one fixed point $q=0$ while for the
ferroelectric case there are three fixed points, the unstable one
$q=0$ and the two stable ones $q=\pm 1$ corresponding to two
opposite signs of the polarization. It is then natural to introduce
the natural frequency of the oscillator $\omega_0^2= m$ for the
paraelectric case and $\omega_0^2= m+3q^2_0$ for the ferroelectric
case. We therefore consider below the general linear problem of a harmonic
oscillator of frequency $\omega_0$ embedded in a cavity.

\section{The linear limit: normal modes}

The linear problem is
\begin{eqnarray}
e_{tt}-e_{xx} &=&-\alpha \delta \left( x-x_a\right)q_{tt},  \label{lineq} \\
q_{tt}+\omega_0^2 q &=&\alpha e(x_a).
\end{eqnarray}%
Note that we have a Dirac delta function in the first equation so that the
solution will not have a second derivative at $x=x_a$. In this case, one can
write the solution using standard sine Fourier
modes. However these are not adapted to describe the evolution because the
projection of the operator gives wrong results \cite{Hildebrand}. Then 
we need to define new
normal modes.
For this one first separates time and space and one looks for solutions
in the form
\[
e(x,t) = E(x)e^{i\omega t},~~ q(x,t) = Q(x)e^{i\omega t}
\]
so that the system (\ref{lineq}) becomes \be
E^{\prime\prime}(x)+\omega ^2E(x)=-\alpha\omega ^2 Q \delta
(x-x_a),~~~ \label{harmeq} Q=\frac{\alpha E(x_a)}{\omega_0^2-\omega
^2}, \ee As expected from the general theory of linear operators
\cite{Courant_Hilbert} the system will exhibit eigenfrequencies and
eigenmodes (normal modes). Combining these two equations, we obtain
the final boundary value problem for $E$
\begin{equation}
E^{\prime \prime }(x)+\omega ^{2}\left( 1+\frac{\alpha ^{2}\delta
(x-x_{a}) }{\omega_0^2-\omega^{2}}\right) E(x)=0 , \label{sderiv2}
\end{equation}
with the boundary conditions $E(0)=E(l)=0$.

To obtain the solution, notice that except for $x=x_a$
$$E^{\prime\prime}(x)+\omega ^2E(x)=0.$$
Using this remark and the boundary conditions we get the
left and right solutions
\begin{equation}\label{sol}
E(x)=\left\{ \begin{array}{l l}
A \sin\omega x,  & x<x_a,\\
B \sin\omega (l-x), & x>x_a,
\end{array}\right.\end{equation}
where $A$ and $B$ are constants.
To connect the left and right solutions we use the
continuity of $E(x)$ as well as of $e$ at
$x=x_a$. The second relation needed is the jump condition for $E'$
obtained by integrating (\ref{sderiv2}) over a small interval
centered on $x_a$. When the size of the interval goes to zero we get
\begin{equation}
[E^{\prime}]_{x_a^{-}}^{x_a^{+}}=-\frac{\alpha^2\omega ^2 }{
\omega_0^2-\omega ^2}E(x_a).  \label{jump}
\end{equation}
At $x=x_a$ the continuity of $E$ and jump condition (\ref{jump}) give the
following relations
\begin{eqnarray}  \label{homog_sys}
A \sin\omega x_a - B \sin\omega (l-x_a) = 0 , \\
A ( {\alpha^2 \omega \over \omega_0^2 -\omega^2}\sin\omega x_a
- \cos\omega x_a )
-B \cos\omega(l-x_a)= 0.
\end{eqnarray}
For this homogeneous system to have a non zero solution, the
determinant must be zero and this gives the dispersion relation 
\be
\label{dispersion} \sin \omega l = \alpha^2 {\omega \over \omega_0^2
-\omega^2} \sin \omega x_a \sin \omega (l-x_a) ,\ee 
which determines the allowed frequencies $\omega$ of the system.

As can be expected from the general theory \cite{Courant_Hilbert}, we 
have an infinite countable set of allowed frequencies.
Note that in the absence of coupling to
the film $\alpha=0$, we get the standard Fourier modes $\omega_n =
{n \pi /l}$. For small $\alpha$ the shift in frequency is small
because the right hand side is proportional to $\alpha^2$. These
eigenfrequencies can be computed using bisection for example.

To summarize, the eigenvalues and
eigenvectors (up to a multiplicative constant) of the boundary value
problem (\ref{harmeq}) are
\begin{equation}\label{V_i}
\left\{\omega_i,~~{\mathbf V_i}=\left(
\begin{array}{c} E_i(x) \\ Q_i\end{array}\right)\right\}.
\end{equation}
where
\begin{equation}\label{ei}
E_i(x)=\left\{ \begin{array}{l l}
\sin \omega_i x,  & x<x_a,\\
{\sin \omega_i x_a \over \sin \omega_i (l-x_a)} \sin\omega_i (l-x), &
x>x_a,
\end{array}\right .~~ ,Q_i = \frac{\alpha E_i(x_a)}{\omega_0^2 -\omega_i^2}
, \end{equation}
and $\omega_i$ satisfies (\ref{dispersion}).

Note that an equation similar to (\ref{sderiv2}) was obtained 
in the theory of a 1D waveguide with a perfect mirror at one
end and a two-level atom at the other end \cite{Dong}. Contrary to our
case, this is not an eigenvalue problem because the system is
open on one side. \\
From another point of view, the system cavity/film (\ref{lineq}) was
considered for $\omega_0=0$ by Bocchieri et al \cite{bcl72} in the context
of statistical mechanics. Their main result was that there 
was always energy exchange
between the film and the cavity so that equipartition could not
be reached. Indeed, this can be seen by examining the normal modes
(\ref{ei}) which couple $E_i$ and $Q_i$. Only for special symetries
(film at the center of the cavity \dots) and special frequencies
do we get $Q_i=0$.

\subsection{Influence of the film parameters on the dispersion relation}

\begin{figure}
\centerline{\epsfig{file=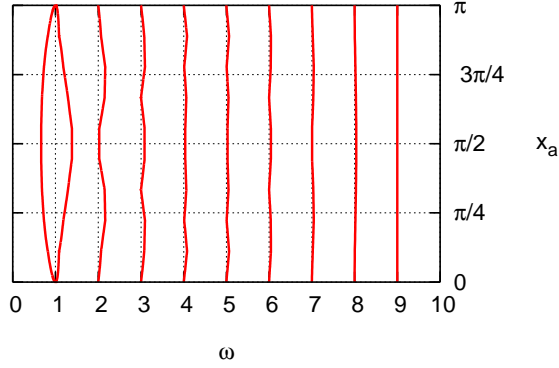,width=0.9\linewidth,angle=0}} 
\caption{Eigenvalues (zeros of the dispersion relation
(\ref{dispersion}) ) as a function of the position of the film
$x_a$. The parameters are $l=\pi$,  $\alpha=1$ and $\omega_0=1$}
\label{f1}
\end{figure}

We now study the dispersion relation in more detail.
In Fig. \ref{f1} we plot the
solutions of (\ref{dispersion}) as a function of the film position
for $\omega_0=1$ and $\alpha=1$. Notice how the systems generates
two eigenvalues in place of the frequency $\omega_0=1$. For large
frequencies $\omega \gg \omega_0$ we recover the standard Fourier
cavity modes $\omega_n = n \pi / l$.
The eigenmodes for the particular case of a centered film shown in
Fig. \ref{f1} contain the even Fourier modes. These correspond to $Q_i=0$
because their derivative is continuous at $x_a$. This will have important
practical consequences.

Because of the film, the eigenfrequencies of the system
film/cavity differ from the usual Fourier cavity modes $\omega_n = n \pi/l$.
They are shifted if $\omega_n \neq \omega_0$ and
disappear if $\omega_n = \omega_0$.
Let us compute this shift in the limit of small $\alpha$ using
perturbation theory.
To simplify the analysis, we assume $l=\pi$ so that $\omega_n= n \pi/l=n$.
We search the frequency $\omega$ using
the expansion
\[
\omega=n+ \alpha^2 \omega_1+..
\]
with $\omega_1 \ll 1$ and $n$ are the usual sine Fourier modes of the cavity.
We have the following relations
$$\sin \omega l = (-1)^n \alpha^2 \omega_1 \pi + O(\alpha^4) $$
$$\sin \omega x_a = \sin n x_a + \alpha^2 \omega_1 x_a \cos n x_a +O(\alpha^4) $$
$$\sin \omega (l-x_a) = \sin n (l-x_a) + \alpha^2 \omega_1 (l-x_a) \cos n (l-x_a) +O(\alpha^4) $$
Plugging these relations into (\ref{dispersion})
we have, up to $O(\alpha^4)$
\begin{equation}
(-1)^n \omega_1 = {n \over \pi(\omega_0^2-n^2)} \sin n x_a \sin n (l-x_a).
\label{dispers4}
\end{equation}
Assuming that $\alpha^2 < 1$ we get the simplified expression
\begin{equation}
\omega = n - \alpha^2 (-1)^n {n\over \pi}{ 1\over \omega_0^2-n^2} \sin n x_a \sin n (l-x_a) 
+ O(\alpha^4) .
\label{dispers5}
\end{equation}
Due to the presence of the film the cavity
modes close to the film mode are blue shifted
if the frequency of cavity mode is above the
oscillator eigenfrequency and red shifted for lower cavity mode
frequencies.

When $\omega_0^2=n_0^2$ where $n_0$ is an integer, the oscillator
frequency coincides with the cavity mode. In this case the
eigenfrequency of the combined system splits away from $n_0$. Again
this can be calculated for small $\alpha$ by assuming
$$\omega = \omega_0 + \omega_1 , $$
where $|\omega_1 | \ll \omega_0$.

Plugging this expression into the dispersion relation and collecting
the terms, we obtain the second degree equation

$$A {\omega_1}^2 + B \omega_1 +C = 0,$$
where
\begin{eqnarray}\label{abc}
A = 4 \pi n +\alpha^2 ( (\pi-2 x_a) \sin2n x_a-(\cos 2nx_a -1)n/2) ,\\
B = \alpha^2 (n(\pi-2 x_a) \sin 2n x_a +\cos 2n x_a
-1) ,\\
C = \alpha^2 n (\cos 2n x_a-1).
\end{eqnarray}

The two branches of the resonant frequency are then given by
$$\omega_1 = {-B \pm \sqrt{B^2-4AC} \over 2A}.$$
As an example, consider the case of Fig \ref{f1} corresponding to
$l= \pi$ and a film placed in the center of the cavity $x_a=0.5 l$.
Then $\cos \omega (l-2x_a)=1$ and  $\cos \omega l\approx
(-1)^{n_0}(1-\omega^2_1/2)$

Then the splitting is given by
\begin{equation}
4 \pi n_0 \omega_1^2
=\alpha^2(n_0+\omega_1)(1-\omega^2_1/2)(1-(-1)^{n_0})
\label{dispersres1}
\end{equation}
so that $\omega_1=0$ for even resonant frequency $n_0$ and there is
no shift from the resonance in this case. For the odd $n_0$ we
obtain the quadratic equation for the frequency detunings $\omega
_1$ from the resonance:
\begin{equation}
\omega^2_1 -\frac{\alpha^2 \omega_1}{n_0(2
\pi+\alpha^2/2)}-\frac{\alpha^2}{2 \pi +\alpha^2/2}=0
\label{dispersres2}
\end{equation}

\begin{equation}
\omega_1 = \frac{\alpha^2}{2 n( 2 \pi + \alpha^2/2)} \left(1 \pm
{\sqrt {1+\frac {4 {n_0}^2(2 \pi + \alpha^2/2)}{\alpha^2} }} \right)
\label{dispersres3}
\end{equation}

\begin{figure}
\centerline{\epsfig{file=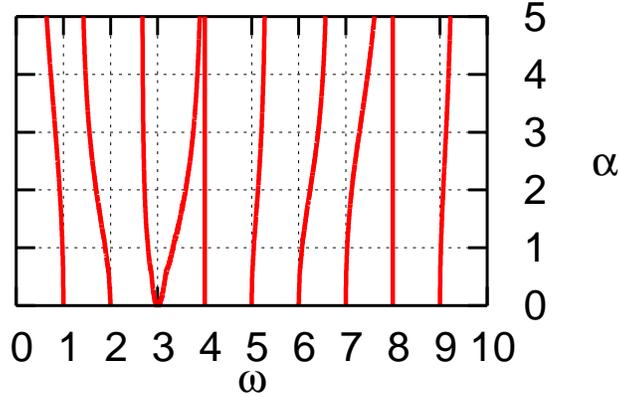,height=0.7\linewidth,angle=-90}}
\caption{Plot of the first 10 eigenfrequencies as a
function of the coupling parameter $\alpha$ for an off-centered film
$x_a=l/4=\pi/4$ whose frequency is $\omega_0=3$.} \label{f1a}
\end{figure}

When $\alpha$ increases, the influence of the film grows and the
eigenfrequencies become very different from the Fourier cavity
modes. In fact when $\alpha \gg 1$, the dominating term in the
dispersion relation is the right hand side and we obtain
$$\sin \omega x_a = 0, ~~~{\rm or}~~\sin \omega (l-x_a)=0.$$
so that \be\label{largea} \omega_n = {n \pi \over x_a}, ~~~{\rm
or}~~\omega_n = {n \pi \over l-x_a}. \ee This corresponds to oscillations
in the left or right cavities. Fig. \ref{f1a} shows the first 10
eigenfrequencies as a function of the coupling parameter $\alpha$
for a cavity of length $l=\pi$ and an off-centered film $x_a=l/4$
whose frequency $\omega_0=3$.
Notice the splitting for $\omega=3$. As $\alpha $ increases, the
eigenfrequencies tend to the ones given by (\ref{largea}), ie 
$\omega_n = 4n$ or $\omega_n = 4n/3$.

\section{Orthogonality of the normal modes }

Using the vector notation $\mathbf V$ defined above, the original
linear system (\ref{lineq}) can be formally written as
\be\label{sys_t0}
(\partial^2_t + {\mathbf L} ) {\mathbf V} = 0 ,\ee
where the operator $\mathbf L$ is
\begin{equation} \label{operator}
\mathbf L=-\partial _{x}^2 \mathcal{\left( \matrix{ 1 & 0 \cr  0 & 0
\cr}  \right)}+\left( \matrix{ \alpha^2 \delta(x-x_a)&
-\alpha \omega_0^2 \delta(x-x_a) \cr  -\alpha\int \delta(x-x_a) & \omega_0^2
  \cr}  \right).
\end{equation}
The eigenfrequencies and eigenvectors $\omega_i,~~{\mathbf V_i}$ are such
that
\be\label{lvi}
L {\mathbf V_i} = \omega_i^2 {\mathbf V_i}.
\ee

The boundary value problem (\ref{sderiv2}) is not a standard Sturm-Liouville
problem because the potential depends on the eigenvalue. Therefore
one needs to define a particular scalar product so that the normal
modes $\mathbf V_i$ defined previously are orthogonal. This is crucial
if we want to use these modes as vectors on which we can project the
state of the linear system (\ref{lineq})
and therefore obtain a simplified description.

To define this scalar product
consider eq. (\ref{harmeq}) with two solutions
\begin{eqnarray} E_j^{\prime \prime}(x)+\omega_j^2 E_j(x)+
\alpha \omega_j^2 Q_j\delta(x-x_a)=0 \\
E_i^{\prime \prime}(x)+\omega_i^2 E_i(x)+ \alpha \omega_i^2
Q_i\delta(x-x_a)=0.
\end{eqnarray}
To show orthogonality the equation for $E_i$ is multiplied by $E_j$
and vice-versa.  Subtracting the resulting equations one obtains:
\begin{equation}
E_iE_j^{\prime \prime}-E_jE_i^{\prime \prime}+ (\omega_j^2
-\omega_i^2) E_iE_j+ \alpha\delta(x-x_a)\left(E_i\omega_j^2 Q_j-E_j
\omega_i^2 Q_i\right)=0 \label{sol1}
\end{equation}
After integration the resulting equation on the domain $[0;l]$ the
first two terms drop out because of the boundary conditions.
Substitute $Q_i, Q_j$ in the last term using (\ref{ei}) leads to the
the following
\begin{equation}\label{rthgnlt}
\left(\omega_j^2 -\omega_i^2\right)
\left[\int_0^l E_iE_jdx + \omega_0^2 Q_iQ_j \right]=0.
\end{equation}
This shows that for $\omega_i\neq \omega_j$ the term in the brackets
should be zero.
The scalar product is then defined as
\begin{equation}\label{sclr prdct}
\left<{\mathbf V_i};{\mathbf V_j}\right>  \equiv
\int_0^l E_iE_jdx + \omega_0^2 Q_iQ_j.
\end{equation}
The relation (\ref{sclr prdct}) establishes a strictly positive
linear form, so it is a scalar product.

When the film has a finite thickness $h$ the Dirac distribution
needs to be replaced by the characteristic function
$$g(x)= 1 ~~{\rm for }~~ |x-x_a| < h/2,~~~0 ~~{\rm elsewhere } .$$
The scalar product becomes
\begin{equation}\label{sclr prdctg}
\left<{\mathbf V_i};{\mathbf V_j}\right>  \equiv
\int_0^l [E_iE_j + \omega_0^2 Q_iQ_j ]g(x)dx.
\end{equation}

Eq.(\ref{rthgnlt}) shows the orthogonality of the eigenvectors $V_i$
for the scalar product defined by eq.(\ref{sclr prdct}). Now it is
possible to choose $A_i$ such that the vectors are normalized
$$\left<{\mathbf V_i};{\mathbf V_i}\right>=1.$$
For this we compute $\left<{\mathbf V_i};{\mathbf V_i}\right>$.
{\small
\begin{eqnarray*}
\left<{\mathbf V_i};{\mathbf V_i}\right>= A_i^2\left(\int_0^a
\sin^2(\omega_i x) dx + \frac{\sin^2(\omega_i x_a)}{\sin^2(\omega_i
(l-x_a))}\int_a^l \sin^2(\omega_i (l-x)) dx + \omega_0^2
\frac{\alpha^2 \sin^2(\omega_i x_a)}{(\omega_0^2
-\omega_i^2)^2}\right),
\\ =
{A_i^2}\left( \frac{x_a}{2}- \frac{\sin 2\omega_i x_a}{4\omega _i}+
\frac{\sin^2 \omega_i x_a}{\sin^2 \omega_i (l-x_{a})} \left( \frac
{l}{2}-\frac {x_a}{2}- \frac {\sin 2 \omega_i
(l-x_a)}{4\omega_i}\right) +\frac{\omega_0^2 \alpha^2\sin^2 \omega_i
x_a}{(\omega_0^2-\omega_i^2)^2}\right)
\end{eqnarray*}
}
Consequently, ${\mathbf V_i}$ is an orthonormal basis when
{\small
\begin{equation}\label{A_i}
A_i = \left( \frac{x_a}{2}- \frac{\sin 2\omega_i x_a}{4\omega _i}+
\frac{\sin^2 \omega_i x_a}{\sin^2 \omega_i (l-x_{a})} \left( \frac
{l}{2}-\frac {x_a}{2}- \frac {\sin 2 \omega_i
(l-x_a)}{4\omega_i}\right) +\frac{\omega_0^2 \alpha^2\sin^2 \omega_i
x_a}{(\omega_0^2-\omega_i^2)^2}\right)^{-1/2}
\end{equation}
}
\begin{figure}
\centerline{
\epsfig{file=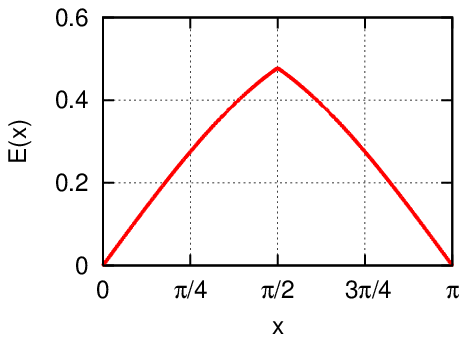,height=5cm,width=5cm,angle=0}
\epsfig{file=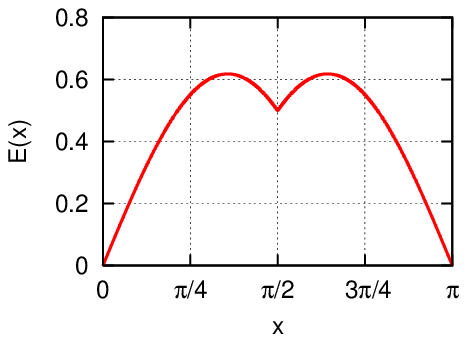,height=5cm,width=5cm,angle=0}
\epsfig{file=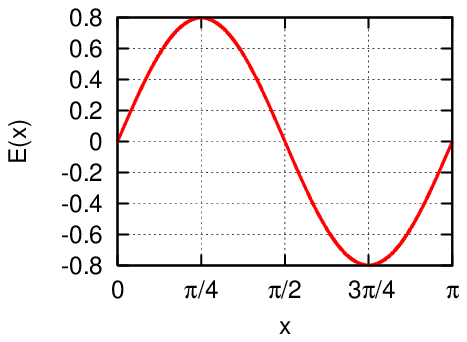,height=5cm,width=5cm,angle=0}
}
\caption{Plot of the $E$ component of the normalized first, second
and third eigenmodes. The film is in the middle of the cavity
$x_a=\pi/2$. The other parameters are the same as in Fig. \ref{f1}.}
\label{f2}
\end{figure}

\begin{figure}
\centerline{
\epsfig{file=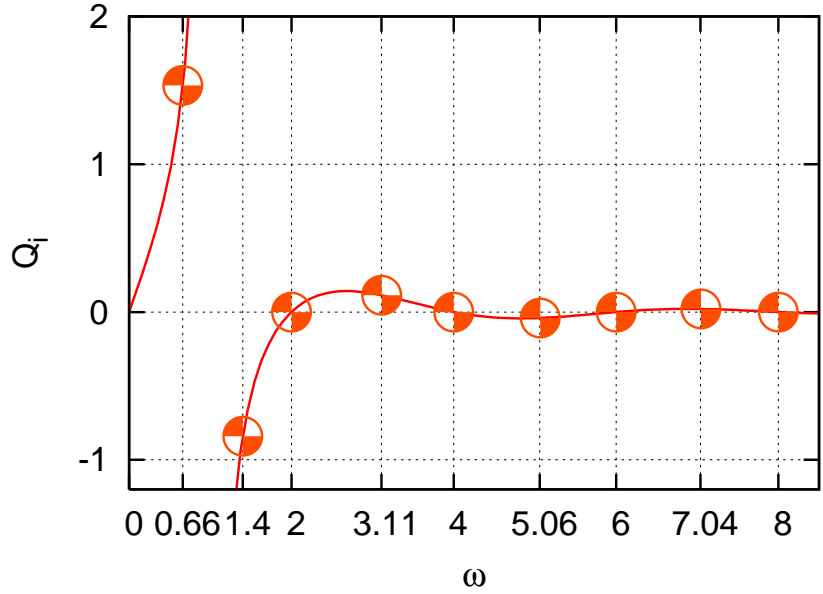,height=0.9\linewidth,angle=-90} 
}
\caption{Plot of the $Q_i$ components of the normalized eigenmodes
as a function of $\omega$ for $x_a=\pi/2 , l=\pi$. The eigenvalues
are indicated by the vertical lines. } \label{f3}
\end{figure}

The normalized eigenvalues are plotted in Fig. \ref{f2} together
with the associated $Q_i$
in Fig. \ref{f3} for a film placed in the center of a cavity
of length $l=\pi$. Notice the clear break in the derivative at $x_a$.
The orthogonality of the modes $(\mathbf V_i, \mathbf V_j)$ comes from the
compensation of the integral of $E_i E_j$ by the product
$Q_i Q_j$.

\begin{figure}
\centerline{ 
\epsfig{file=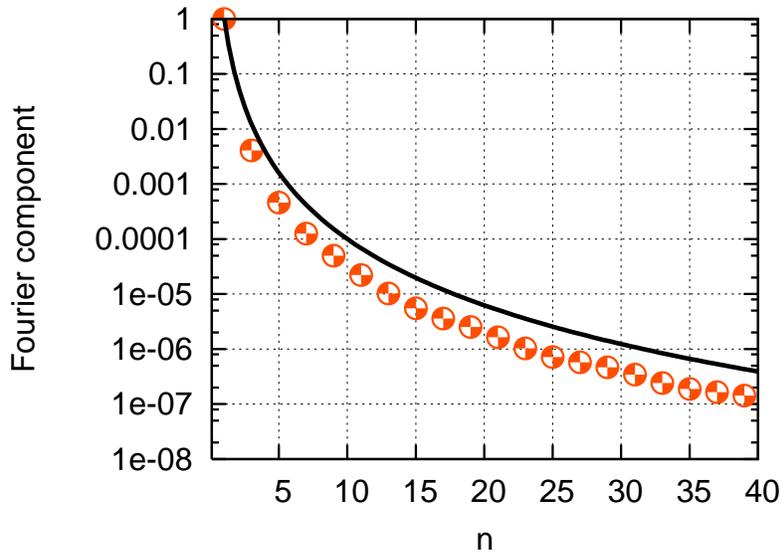,height=0.9\linewidth,angle=-90} 
}
\caption{Plot of the squares of the Fourier sine
coefficients of the $E$ component of the first normal mode in
log-linear scale. The function $1/n^4$ is plotted in continuous
line.} \label{f3.1}
\end{figure}

These modes are specially adapted to describe the coupled system
film/cavity. Many standard sine Fourier modes are necessary to
get a good approximation of the first normal mode. This is seen
in Fig. \ref{f3.1} which shows the amplitude square of the sine
Fourier coefficients of the first normal mode. Notice the typical
$1/n^4$ decay due to the fact that the second derivative of $E$
is singular at $x=x_a$ \cite{Carslaw}.

\begin{figure}
\centerline{
\epsfig{file=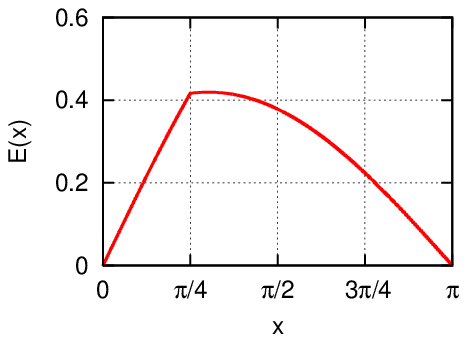,height=5cm,width=5cm,angle=0}
\epsfig{file=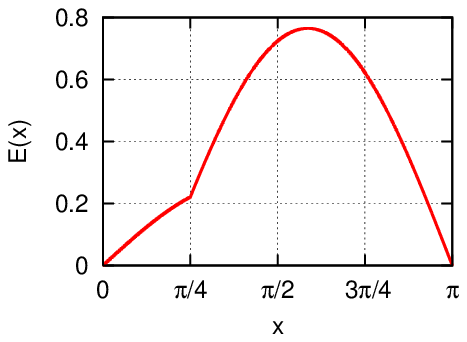,height=5cm,width=5cm,angle=0}
\epsfig{file=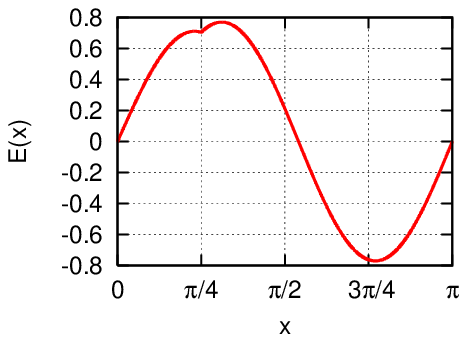,height=5cm,width=5cm,angle=0}
}
\caption{Plot of the $E$ component of the normalized first, second
and third eigenmodes. The film is shifted to the left of the
cavity
$x_a=\pi/4$. The other parameters are the same as in Fig. \ref{f1}.}
\label{f4}
\end{figure}

\begin{figure}
\centerline{
\epsfig{file=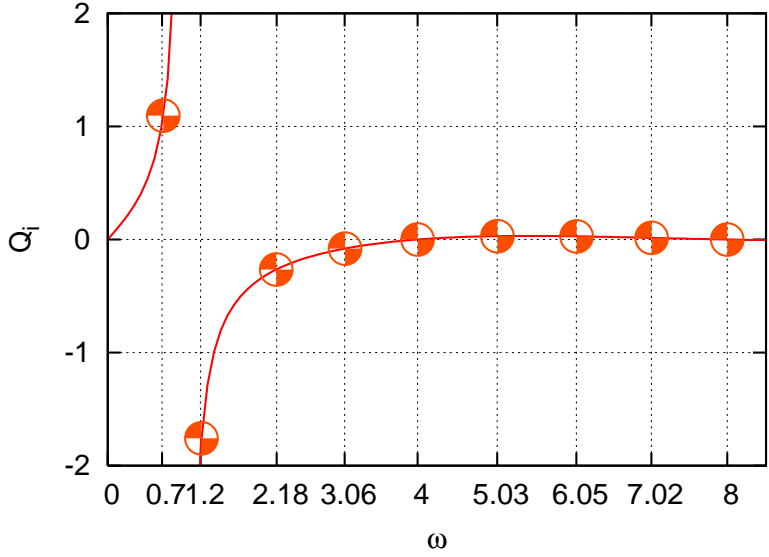,height=0.9\linewidth,angle=-90} 
}
\caption{Plot of the $Q_i$ components of the normalized eigenmodes
as a function of $\omega$ for $x_a=l/4,~l=\pi$ . The eigenvalues are
indicated by the vertical lines. } \label{f5}
\end{figure}

When the film is shifted to one side of the cavity, the modes become
asymmetric as shown in Fig. \ref{f4}. Again the break in the
derivative is clearly apparent. Here standard Fourier modes only
appear for $n=4,8,..$. The $Q_i$ decay very quickly to 0 as shown in
Fig. \ref{f5}.

\section{Cavity mode transfer using an active film}

The normal modes defined in the previous section define
a basis to describe the state of the combined system
cavity/film. We now show that it is possible by acting on the
film to switch the cavity from one mode to another neighboring
mode. This feature is impossible for a single linear system.
It is possible here because of the combination of the two
linear subsystems: the cavity and the film.

In order to describe analytically this process, we introduce
the forcing of the film as $f(q,q_t,t)$ and write the system
as
\begin{eqnarray}
e_{tt}-e_{xx} &=&-\alpha \delta \left( x-x_a\right)q_{tt},  \label{forc0} \\
q_{tt}+\omega_0^2 q &=&\alpha e (x_a)+f(q_t,t),
\end{eqnarray}%

Using the vector notation $\mathbf V$ of
the previous section, this system can be
formally written as
\be\label{sys_t}
(\partial^2_t + {\mathbf L} ) {\mathbf V} = {\mathbf F} ,\ee
where the operator $\mathbf L$ is given by (\ref{operator})
and the forcing is
\be \mathbf F = \left(
\begin{array}{c} - \alpha \delta(x-x_a) f  \\ f \end{array}\right).\ee

For this linear system, it is natural to expand the
state vector $\mathbf V$ in terms of the (normalized) normal modes
\begin{equation} \label{aivi}
\mathbf V=\left( \matrix{ e \cr q \cr } \right)=\sum_i \alpha_i (t)
\mathbf V_i ,
\end{equation}
where the normal modes $\mathbf V_i$ verify the relation
${\mathbf L} {\mathbf V_i} = \omega_i^2 {\mathbf V_i}$.
Plugging (\ref{aivi}) into the equation (\ref{sys_t}) and projecting over
each normal mode we get
\be\label{ait}
{\ddot \alpha_i}  +\alpha_i \omega^2_i =<\mathbf V_i \mathbf F>,~~~i=1,2, \dots
\end{equation}
where the scalar product
\begin{equation}
<\mathbf V_i \mathbf F > = \int_0^l E_i (-\alpha f \delta
(x-x_a))dx+\omega^2_0 Q_i f = \omega_i^2 Q_i f .\label{forc3}
\end{equation}

We now assume that $f$ consists in a damping and forcing term
\be\label{force}
f(t)=-\gamma (t)q_t + I(t) .
\ee
Recalling the linear combination (\ref{aivi}) $q_t = \sum_n {\dot \alpha_n} Q_n$
we get the final expression of the scalar product (\ref{forc3})
\begin{equation}
<\mathbf V_i \mathbf F > = -\gamma (t) \omega^2_i Q_i \sum_n {\dot \alpha_{n}}
Q_n + \omega^2_i Q_i I(t). \label{forc6}
\end{equation}

To illlustrate this, consider just two modes in the expansion (\ref{aivi}).
The system describing the evolution of the mode amplitudes is then
\begin{eqnarray}\label{ampli_eq}
{\ddot \alpha_1}  +\alpha_1 \omega^2_1 & = & -\gamma (t) \omega^2_1 Q_1 (
{\dot \alpha_{1}}Q_1 +{\dot \alpha_{2}}Q_2) + \omega^2_1 Q_1 I(t), \\
{\ddot \alpha_2}  +\alpha_2 \omega^2_2 & = & -\gamma (t) \omega^2_2 Q_2 (
{\dot \alpha_{1}}Q_1 +{\dot \alpha_{2}}Q_2) + \omega^2_2 Q_2 I(t)
\end{eqnarray}
Notice that only using the normal modes (\ref{V_i},\ref{ei}) and
the scalar product (\ref{sclr prdct}) does one obtain a consistent
modal description of the system. Using for example the standard
Sturm-Liouville modes associated to a linear impurity placed at
$x=x_a$ leads to an inconsistency. This important
fact is shown in appendix A. \\
Also remark that for $l=\pi,~a=l/2$, the normal modes include the
even (standard) sine Fourier modes. These however are decoupled from
all the other modes because for them $Q_i=0$ so the right hand side
of the amplitude equation (\ref{ampli_eq}) is zero.

\section{Numerical simulations }

To test these ideas, we have undertaken numerical simulations of the
equations (\ref{lineq}) using the method of lines, where the spatial
operator is integrated over reference intervals (finite volume method).
The time evolution is then done using an ordinary differential equation
solver. The algorithm is described in appendix B.

\subsection{Linear regime}

We introduce a characteristic forcing time function \be\label{goft}
g(t) = {1 \over 2}\left[\tanh\left({t-t_1 \over
w_t}\right)-\tanh\left({t-t_1 \over w_t}\right) \right]\ee and
assume that the damping and forcing are \be\label{damp_force}
\gamma(t) = g(t) \gamma_0 ,~~~I(t) = g(t) \beta \sin(\omega t
+\phi),\ee where $\gamma_0, \beta,\omega$ and $\phi$ are parameters.
We consider the case of a centered film $x_a=l/2 = \pi/2$.
For all the runs presented, we chose $\gamma_0=0.1, \beta=0.2, \phi=0$
and a time interval of forcing $[t_1,t_2]=[50,100]$ with
$w_t=1$. Unless otherwise specified, the initial state of the system is the
first normal mode.

\begin{figure}
\centerline{
\epsfig{file=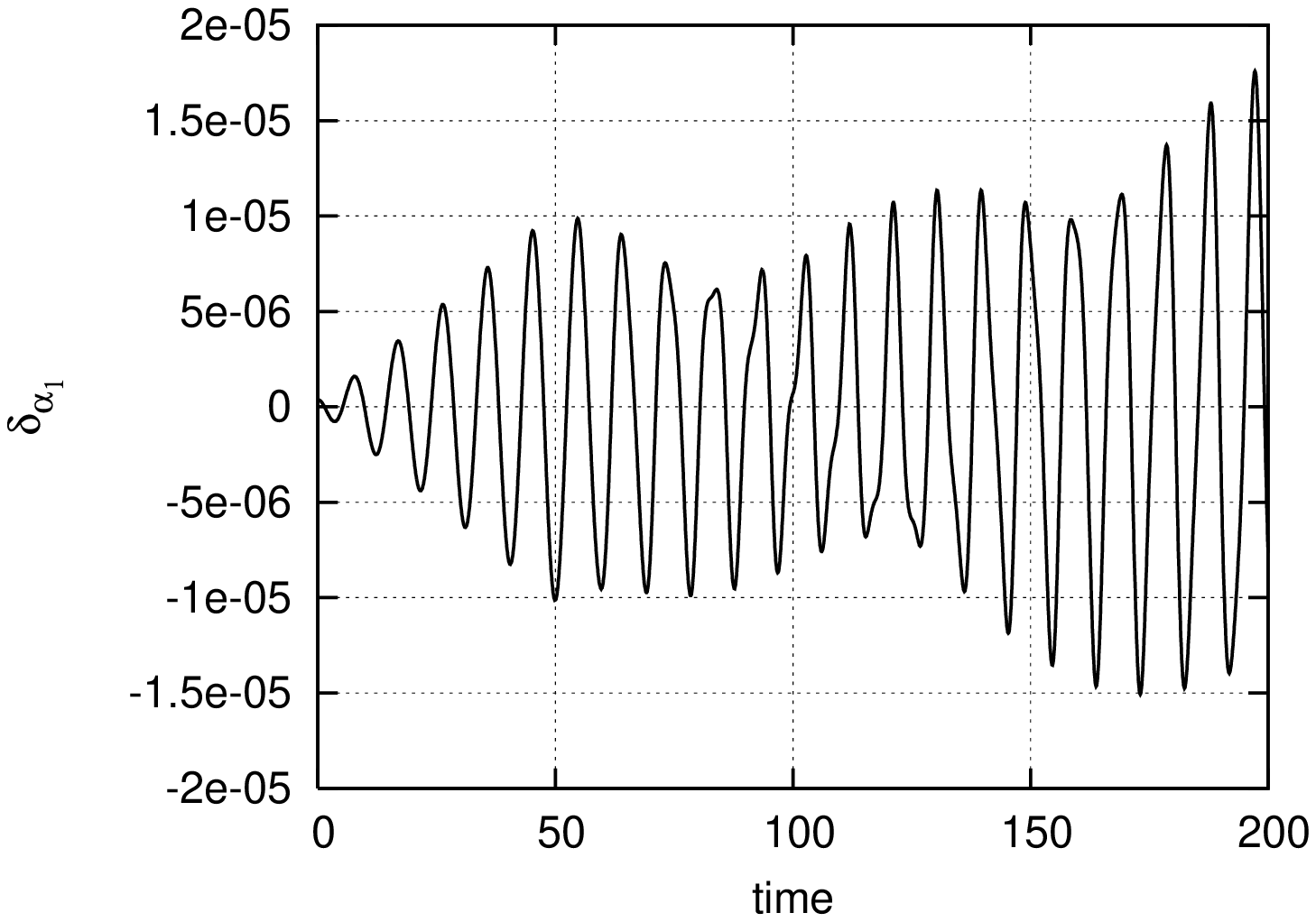,height=0.35\linewidth,angle=-90}
\epsfig{file=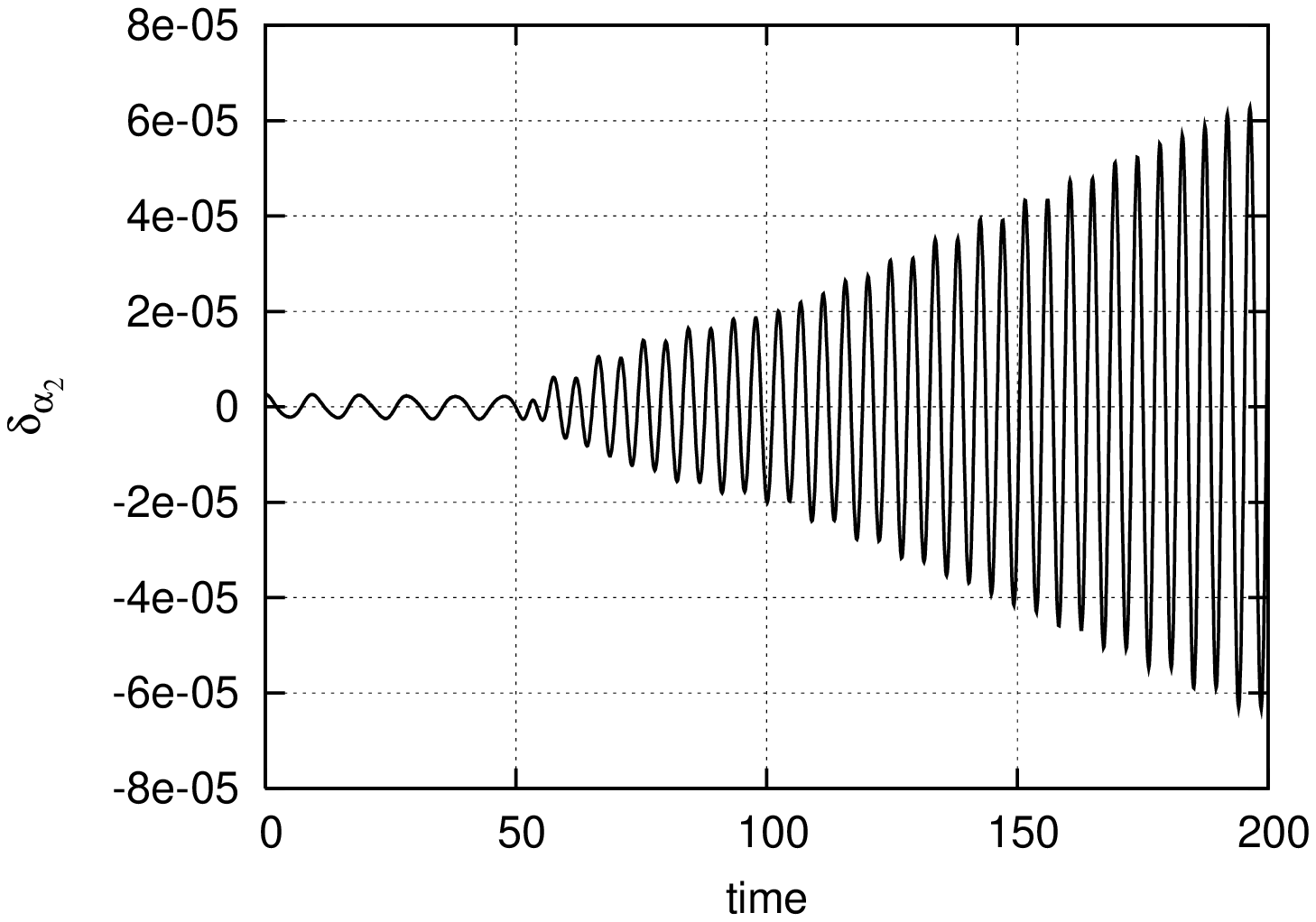,height=0.35\linewidth,angle=-90}
\epsfig{file=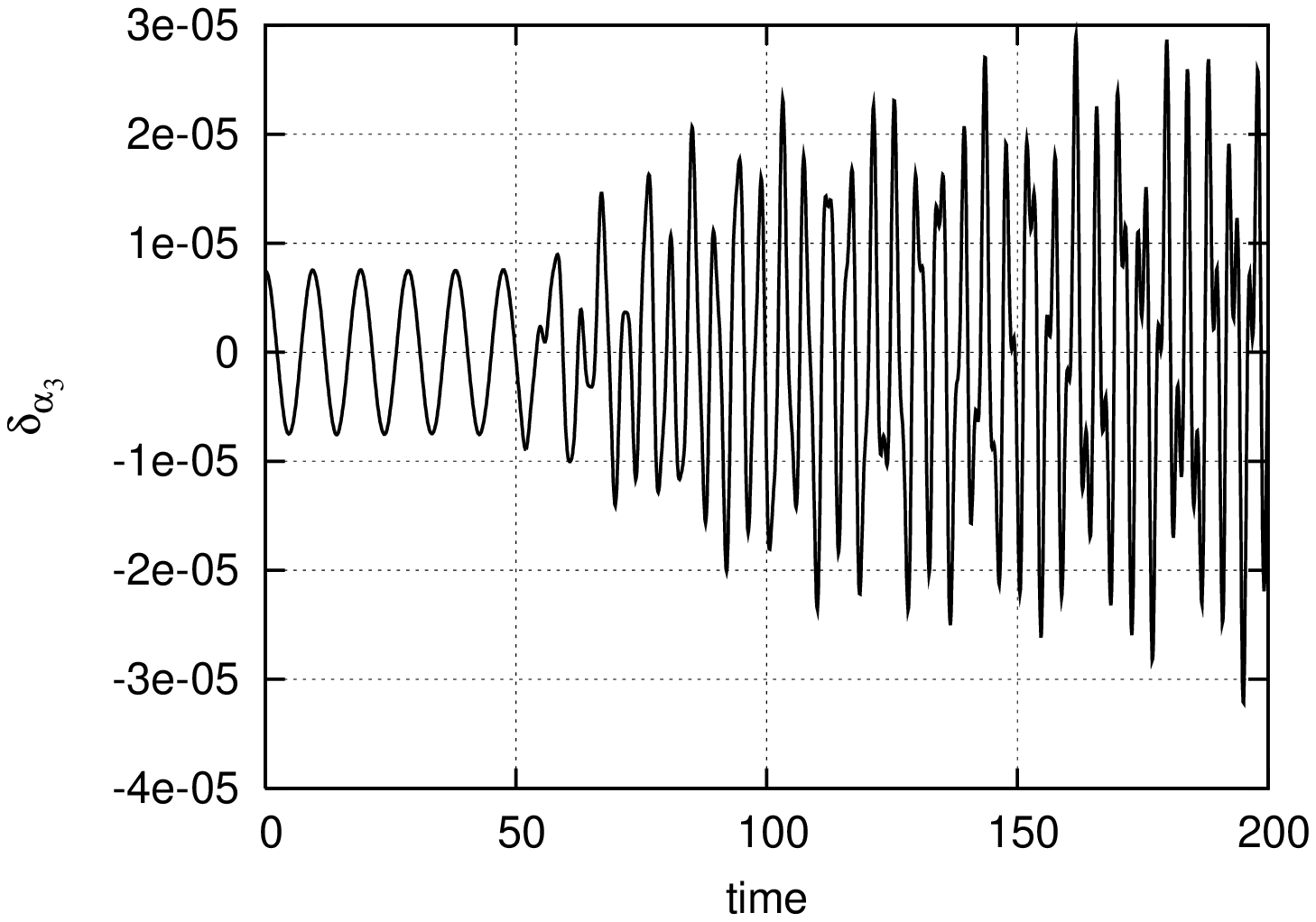,height=0.35\linewidth,angle=-90}
}
\caption{Time evolution of the absolute 
difference $|\alpha_i(t)-\alpha_i'(t)|, ~i=1,2,3$ (from left to right)
between
the solution of the amplitude equations (\ref{ampli_eq}) 
and the coefficients obtained by
projecting the solution of the full partial differential equation
(\ref{lineq}).
The system has been forced at frequency $\omega=1.4\approx \omega_2$
during the time $50<t<100$. }
\label{f5a}
\end{figure}

As a first step, we validate our mode amplitude differential equations
(\ref{ampli_eq}) by comparing their solution with the mode amplitudes
obtained by projecting the solution of the partial differential
equation (\ref{lineq}) onto the normal modes ${\mathbf V_i}$ given by
(\ref{V_i},\ref{ei}). The integrals are calculated using the trapeze
method using 800 mesh points. Fig. \ref{f5a} shows the absolute
error in log scale as a function of time for the first 3 modes
(except even Fourier modes). The difference is consistent with the
error made in the trapeze integration method $O(h^2)$. Notice that
for $i=2$ the difference increases during the forcing. This is due to
the appearance of new modes as shown below.
The agreement is excellent and the 
error of about $10^{-5}$ is essentially the error in the trapeze
method $h^2 = (\pi/800)^2 \approx 10^{-5}$. If only the first three
modes are used in the amplitude equations, the error is still 
very small. 
In many other cases, we compared the
solution of the full problem (\ref{lineq}) with the one given
by the amplitude equations (\ref{ampli_eq}) and always found
errors of about $10^{-5}$. This shows that these simple
amplitude equations are a precise way to describe the complex system
film/cavity.

\begin{figure}
\centerline{
\epsfig{file=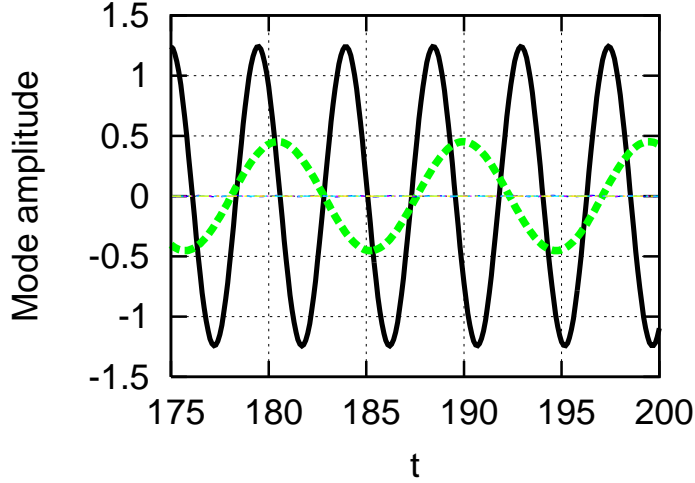,height=0.8\linewidth,angle=-90}
}
\caption{Plot of $\alpha_i(t)$ for $i=1$ (green online) and $i=2$ (continuous 
line) after having forced
the system at frequency $\omega=1.4\approx \omega_2$. The system is 
started in the mode 1 only, $\alpha_1=1$. }
\label{f6}
\end{figure}

\begin{figure}
\centerline{ \epsfig{file=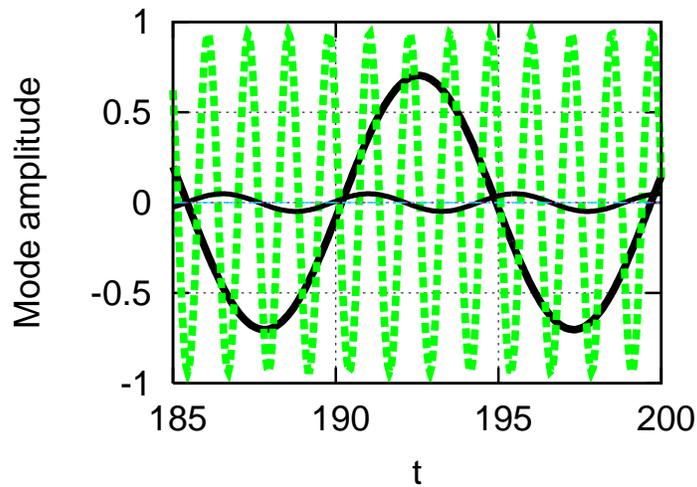,height=0.8\linewidth,angle=-90} }
\caption{Plot of $\alpha_i(t)$ for $i=6$ (dashed line) and $i=1,2$
(continuous line) after having forced
the system at frequency $\omega \approx \omega_1=0.66$. The mode 3
is also present as shown by the blue line. The
system is started in the mode 6 only $\alpha_6=1$. }
\label{f8}
\end{figure}

After this validation, we examine the role of the forcing frequency
and show that we can transfer energy from one cavity mode to another
by acting on the film via the forcing and damping (\ref{damp_force}).
The value of the forcing frequency is essential as shown in the
following pictures. First we chose $\omega=1.4\approx \omega_2$
so that we are
forcing the system to resonate in the second normal mode. The
energy transfer is then efficient and after a short time of
forcing we find the system in the normal mode 2 with very little left of
the normal mode 1. This is shown in Fig. \ref{f6} where we plot the
initial mode in dashed line and the newly generated mode in 
continuous line. This notation will be used throughout this section.
We now change the forcing frequency 
to $\omega=1$ and retain otherwise the same protocol. In this case
we do not have a resonance of the system and it responds by generating
components on the neighboring normal modes. 
When forcing the system on the sine Fourier mode $\omega_3=2$ for which
$Q_3=0$ no energy is fed into this mode as expected from the
amplitude equations (\ref{ampli_eq}). The
system responds on the 1st, 2nd and 4th normal modes.
\begin{figure}
\centerline{ \epsfig{file=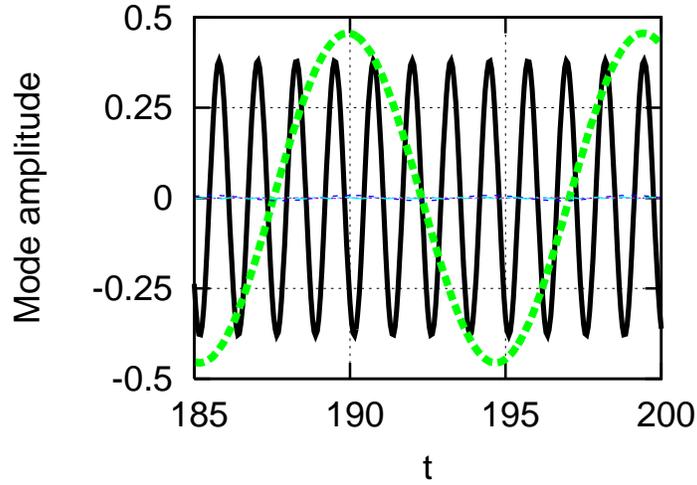,height=0.8\linewidth,angle=-90} }
\caption{Plot of $\alpha_i(t)$ for $i=6$ (continuous line) and $i=1$ 
(dashed line) after having forced
the system at frequency $\omega \approx \omega_6=5.06...$. The 
system is started in the mode 1 only $\alpha_1=1$. }
\label{f7}
\end{figure}
We have forced the system at frequency $\omega=5$ and obtained conversion
from mode 1 to mode 6. This is shown in Fig. \ref{f7}. If we choose $\omega$ closer to $\omega_6$ the 
transfer is even better so that the amplitude of the mode 6 is larger.
Note that it is also possible to obtain down conversion of modes. For
example starting with mode 6 and forcing at a frequency $\omega=0.66
\approx\omega_1$ we obtain the mode 1 and a little of mode 2. This 
is shown in Fig. \ref{f8}.
The value of $Q_i$ determines the efficiency of conversion to
or from the mode $i$. For example with a cavity of length $l=\pi$
and a film placed at $a=l/2$, we find $Q_1=1.8$ and $Q_6=-0.15$.
We can then state that in general, conversion from
modes close to $\omega_0$ to normal
modes far from $\omega_0$ is more efficient than the converse.
This is because in the amplitude
equations (\ref{ampli_eq}) the damping of $\alpha_i$ is
proportional to $Q_i^2$ while the
amplification term is
proportional to $Q_i$. So a mode close to resonance with
a large $Q_i$ is damped more than another normal mode with a
smaller $Q_j$. This is shown in Fig. \ref{f7} and \ref{f8}.

To summarize we have shown that this linear system can convert
energy from one normal mode to another. This was thought impossible for a
linear system because of the orthogonality of the normal modes.
Here because we act on the sub-system we are able to do this transfer.
Another important result is the excellent agreement between the
solution of the amplitude equations and the solutions of the initial
problem. This simple method could then be used in practice
to solve the singular partial differential equation (\ref{lineq}). 

\subsection{Nonlinear regime}

Another way to convert energy from one mode to another is through
nonlinearity. A well known example is the famous study of Fermi, 
Pasta and Ulam (see for example the entry in \cite{scott05}) showing
energy recurrence between Fourier modes in a chain of anharmonic oscillators.
Now we consider the film to follow a law with a cubic
nonlinearity and take out the driving and damping terms.

The film equation now incorporates a cubic nonlinearity so that the
composite cavity/film is described by the system (\ref{eq1}).
The cubic term can be treated as in the previous section
and incorporated into the $F$ term of equation (\ref{sys_t}). The
scalar product is
\begin{equation}
\left <\mathbf V_i \mathbf F \right > = \omega^2_i Q_i q^3=\omega^2_i Q_i 
\left (\sum_n {\alpha_{n}} Q_n \right )^3. \label{sp_nlin}
\end{equation}
Then the amplitude equations are
\begin{eqnarray}\label{ampli_eq2}
{\ddot \alpha_1}  +\alpha_1 \omega^2_1 & = & \omega^2_1 Q_1 
(\sum_n {\alpha_{n}} Q_n )^3, \\
{\ddot \alpha_2}  +\alpha_2 \omega^2_2 & = & \omega^2_2 Q_2 
(\sum_n {\alpha_{n}} Q_n )^3, \\
\dots \nonumber
\end{eqnarray}
If there is in addition forcing and damping on the system, one
needs to add to the right hand side of (\ref{ampli_eq2}) the
terms on the right hand side of (\ref{ampli_eq}).

As the amplitude of the film polarization $q$ increases, one expects that 
higher and lower frequency normal modes will be excited. This coupling 
to the other modes is clear from the right hand side of the 
amplitude equations (\ref{ampli_eq2}). Of course one should not
increase too much the amplitude of the forcing because then the
wave length of the cavity modes would reduce and become 
comparable with the film thickness. Then approximating the film
by a Dirac distribution would not make sense.

\begin{figure}
\centerline{ \epsfig{file=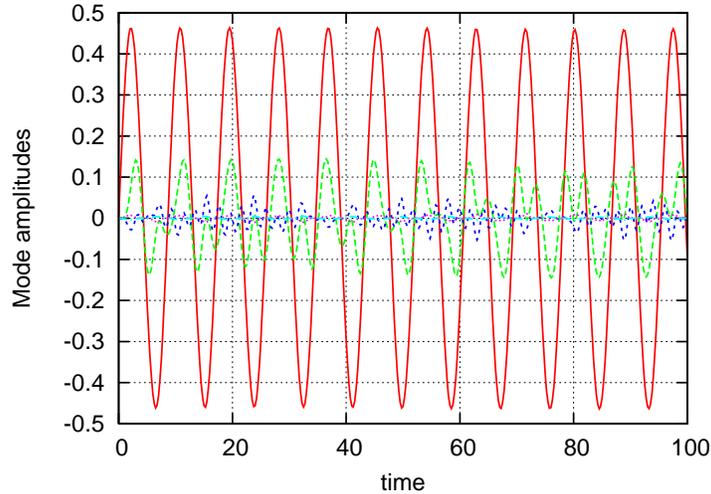,width=0.8\linewidth,angle=0} }
\caption{Plot of $\alpha_i(t)$ for $i=1$ (continuous line), $i=2$ (long dash),
$i=3$ (short dash) for a system with cubic nonlinearity. The results from
the amplitude equations (\ref{ampli_eq2}) with 8 modes are also 
plotted and they superpose exactly. The initial velocity  is 0.5.
The system is started in the mode 1 only. }
\label{f9}
\end{figure}

We start the system in the first normal mode with zero amplitude and
a positive velocity and let it evolve from 0. This procedure is chosen so
as not to create a shock in the system with the nonlinearity.
For small velocities, there is little transfer from the first to
the second and third modes. Again the sine Fourier modes do not
play any role because for them $Q_i=0$. Increasing the initial velocity
increases the transfer of energy to the higher modes. For an velocity
of 0.5, Fig. \ref{f9} shows the evolution of $\alpha_i,~i=1-4$ as a function
of time for both the full system (\ref{nlineq}) and the amplitude 
equations (\ref{ampli_eq2}). 
As expected the nonlinearity generates higher frequencies.
Notice the excellent agreement between the solution of the partial
differential equation system and the amplitude equations. This holds
even for such large velocities as 2.

\begin{figure}
\centerline{ 
\epsfig{file=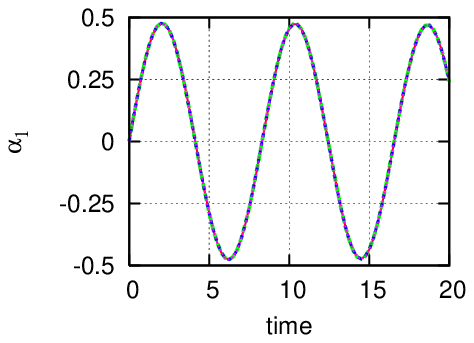,width=0.3\linewidth,angle=0} 
\epsfig{file=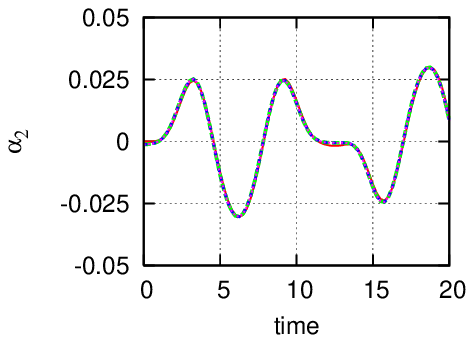,width=0.3\linewidth,angle=0} 
\epsfig{file=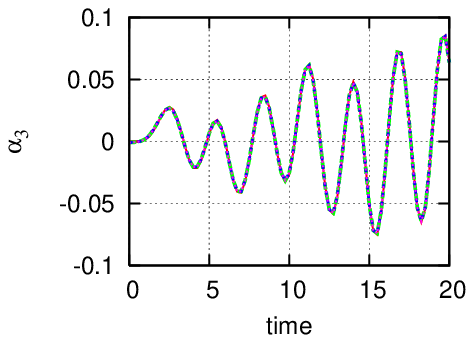,width=0.3\linewidth,angle=0} 
}
\centerline{  
\epsfig{file=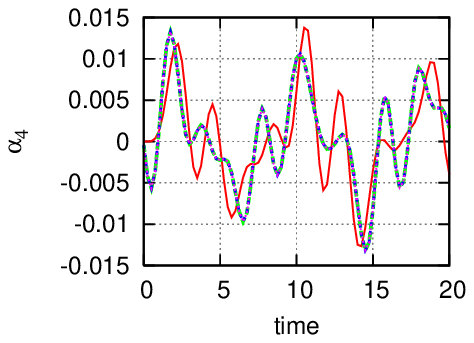,width=0.3\linewidth,angle=0}         
\epsfig{file=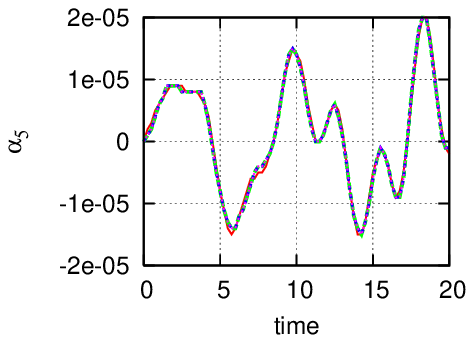,width=0.3\linewidth,angle=0}
\epsfig{file=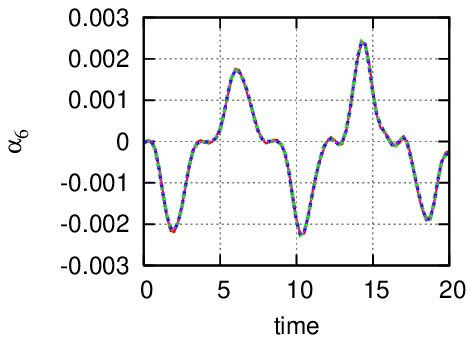,width=0.3\linewidth,angle=0}
}
\caption{Short time evolution of $\alpha_i(t)$ for $i=1-6$ from top 
left to bottom right. The same time evolutions for 
$t= k t_{\rm rec}$ where $t_{\rm rec}=290$ and $k=1-4$ are plotted
on the same panels showing recurrence. The initial velocity  is 0.5
and the system is started in the mode 1. }
\label{f10}
\end{figure}

We conclude this section with an observation of recurrence similar
to what happens for the Fermi-Pasta-Ulam system. Fig. \ref{f10}
presents a short time evolution of the first 6 modes for an initial 
velocity of 0.5, starting from the first mode. In the plots of 
Fig. \ref{f10} are superposed 4 other time evolutions taken at times
$t = k t_{\rm rec}$ where $t_{\rm rec}=290$ and $k$ is an integer.
This recurrence could indicate that our system is close to being
integrable.

\section{Conclusion}

We considered the interaction between an electromagnetic field in a cavity
and a thin polarized dielectric film. The model is the Maxwell-Lorenz
system where the medium is described by an oscillator and the
coupling to the wave equation is through a Dirac delta function.
We introduced normal modes which are adapted to the system film/cavity.
These are well adapted to describe the time evolution of the system,
unlike the standard sine Fourier modes or other Sturm Liouville
modes. The normal modes composed of the field $E$ and the displacement $Q$
are orthogonal with respect to a special scalar product which we
introduce.
The amplitude equations derived from the normal modes give an excellent
description of the dynamics and could even be used as a numerical tool
instead of solving the full partial differential equation system using
the fairly involved finite volume method.

Assuming a linear oscillator for the film, we show conversion from
one normal mode to another by forcing the film at specific frequencies.
This is new for linear systems and could be used for
many applications in optics or microwaves.

If the film is described by an anharmonic oscillator, the evolution
generates other modes. Again the amplitude equations provide excellent 
agreement with the solution of the full problem. Finally we observed 
recurrence for certain initial velocities of the film. This phenomenon
is known to exist for systems close to integrability. The fact that
we observe it here may indicate that our system is in some ways close
to integrability.

{\bf Acknowledgments}

The authors thank Andr\'e Draux and Yuri Gaididei for very useful
discussions. Elena Kazantseva thanks the Region Haute-Normandie for
a Post-doctoral grant. Andrei Maimistov is grateful to the
\textit{Laboratoire de Math\'ematiques, INSA de Rouen} for
hospitality and support. The authors thank the Centre de Ressources
Informatiques de Haute-Normandie for access to computing ressources.

\subsection{Appendix A: Inconsistent projection using standard eigenmodes}

Using standard Sturm-Liouville eigenmodes and the usual scalar product
leads to inconsistent results. To show this let us assume no
forcing for simplicity. We consider the usual
eigenmodes associated with the Sturm Liouville problem
\be\label{stand_sl}
E_{xx} -\alpha^2 E \delta(x-x_a) = -\omega^2 E.\ee
Call these modes $E_n$ associated to the eigenfrequency $\omega_n$.

One then expands the field as
$$\label{e_anen} e(x,t)= \sum_{n=1}^\infty \alpha_n(t) E_n(x).$$
Plugging this expansion into the system of equations (\ref{lineq})
and projecting onto the $E_n$ using the standard scalar product,
one gets the evolution of the $\alpha$'s
\begin{eqnarray}
{\ddot \alpha_1} + \omega_1^2 \alpha_1 = \alpha \omega_0^2 q E_1(a) ,\\
{\ddot \alpha_2} + \omega_2^2 \alpha_2 = \alpha \omega_0^2 q E_2(a) ,\\
{\ddot q} + \omega_0^2 q = \alpha \sum \alpha_i E_i(a).
\end{eqnarray}
Let us examine the jump condition on $E_x$ from the partial differential
equation (\ref{lineq}). We have
\be\label{jump_pde}
-[E_x]_{x_a^-}^{x_a^+} = -\alpha [ -\omega_0^2 q 
+ \alpha E(a) ].\ee
From the expansion (\ref{e_anen}), we obtain
$$-[E_x]_{x_a^-}^{x_a^+}= -\sum_n \alpha_n[{E_n}_x]_{x_a^-}^{x_a^+} .$$
We get the jumps of the ${E_n}_x$ from the
eigenvalue relation
$${E_n}_{xx} -\alpha^2 E_n \delta(x-x_a) = -\omega^2 E_n,$$
and obtain that
$$-\sum_n \alpha_n[{E_n}_x]_{x_a^-}^{x_a^+} = -\alpha^2 \sum_n \alpha_n E_n(a) = -\alpha^2 E(a),$$
which is clearly inconsistent with the result obtained 
from the original system (\ref{jump_pde}). Therefore one needs to use the normal modes
associated with the full system and the special scalar product
(\ref{sclr prdct}) to get
a consistent reduced description of the system.

\subsection{Appendix B: Numerical method for solving (\ref{lineq})}

The evolution equation (\ref{lineq}) involves a partial
differential equation with a Dirac distribution coupled
to an ordinary differential equation. We solve this coupled 
system using the method of lines where the time operator
is kept as such and the space operator is discretized, naturally
leading to a system of coupled ordinary differential equations.
The spatial operator is a distribution so the natural way to give it meaning
is to integrate it over small reference intervals (finite volume 
approximation). The value of the function is assumed to be 
constant in each volume. This method of lines allows to increase
the precision of the approximation in time and space independantly.

We tranform (\ref{lineq}) into a system of first order partial
differential equations
\begin{eqnarray}
\label{edp1} 
e_t  =  f, \\
f_t  =  e_{xx} -\alpha \delta(x-x_a)r_t ,\\
q_t =r ,\\
r_t = -\omega_0^2 q + \alpha e(x_a). \end{eqnarray}

We then define our volume elements making sure that $x_a$
is the center of one element. We then integrate the operator
on each volume $[x_n -h/2,x_n+h/2]$ where $h$ is the space
step. For $x_n \neq x_a$, we recover the standard finite difference
expression for the Laplacian
$$e_{xx} = {e_{n+1}+e_{n+1}-2e_n \over h^2} +O(h^2),$$
and get the discrete wave equation
\begin{eqnarray}\label{dwave}
\dot e_n  =  f_n \\
\dot f_n  =  {e_{n+1}+e_{n+1}-2e_n \over h^2}.\end{eqnarray}
For $x_n = x_a$, we get
\begin{eqnarray}\label{dwaveq}
\dot e_n  =  f_n \\
\dot f_n  =  {e_{n+1}+e_{n+1}-2e_n \over h^2} -\alpha \dot r\\
\dot q =r ,\\
\dot r = -\omega_0^2 q + \alpha e_n . \end{eqnarray}
The coupled system (\ref{dwave},\ref{dwaveq}) is then integrated
using an ordinary differential equation solver.
In practice we use the variable step Runge-Kutta 4-5 Dopri5
developped by Hairer and Norsett at the University of Geneva \cite{hairer}.

\end{document}